\documentclass[aps,prd,onecolumn,superscriptaddress,nofootinbib,showpacs]{revtex4}
\usepackage{epsfig,amsfonts,amsthm}
\usepackage{amsmath,amssymb}
\usepackage{graphicx,color}

\newcommand{\be}{\begin{equation}}
\newcommand{\ee}{\end{equation}}
\newcommand{\bea}{\begin{eqnarray}}
\newcommand{\eea}{\end{eqnarray}}

\renewcommand{\Re}{\mathrm{Re}\,}
\renewcommand{\Im}{\mathrm{Im}\,}
\newcommand{\doublet}[2]{ \left( \begin{array}{c}#1 \\ #2 \end{array}\right) }

\newcommand{\lr}[1]{ \langle #1 \rangle}
\newcommand{\Tr}{\mathrm{Tr}}
\newcommand{\Z}{\mathbb{Z}}
\newcommand{\mmatrix}[4]{ \left(\! \begin{array}{ccc}#1 & #2 \\ #3 & #4 \end{array}\!\right) }
\newcommand{\mmmatrix}[9]{ \left(\! \begin{array}{ccc}#1 & #2 & #3\\ #4 & #5 & #6\\ #7 & #8 & #9\\ \end{array}\!\right) }
\newcommand{\toCP}{\xrightarrow{CP}}
\newcommand{\toC}{\xrightarrow{C}}
\newcommand{\toP}{\xrightarrow{P}}
\newcommand{\bx}{{\bf x}}
\newcommand{\bp}{{\bf p}}

\def\lsim{\mathrel{\rlap{\lower4pt\hbox{\hskip1pt$\sim$}}
    \raise1pt\hbox{$<$}}}         
\def\gsim{\mathrel{\rlap{\lower4pt\hbox{\hskip1pt$\sim$}}
    \raise1pt\hbox{$>$}}}         

\begin{document}

\title{When the $C$ in $CP$ does not matter: anatomy of order-4 $CP$ eigenstates and their Yukawa interactions}

\author{Alfredo Aranda}\thanks{E-mail: fefo@ucol.mx} 
\affiliation{Facultad de Ciencias --- CUICBAS, Universidad de Colima, C.P. 28045, Colima, M\'exico}
\affiliation{Dual CP Institute of High Energy Physics, C.P. 28045, Colima, M\'exico}
\author{Igor P. Ivanov}\thanks{E-mail: igor.ivanov@tecnico.ulisboa.pt}
\affiliation{CFTP, Instituto Superior T\'ecnico, Universidade de Lisboa, Avenida Rovisco Pais 1, 1049--001 Lisboa, Portugal}
\author{Enrique Jim\'enez}\thanks{E-mail: physieira@gmail.com}
\affiliation{Facultad de Ciencias --- CUICBAS, Universidad de Colima, C.P. 28045, Colima, M\'exico}
\affiliation{Dual CP Institute of High Energy Physics, C.P. 28045, Colima, M\'exico}



\begin{abstract}
We explore the origin and Yukawa interactions of the scalars with peculiar $CP$-properties 
which were recently found in a multi-Higgs model based on an order-4 $CP$ symmetry.
We relate the existence of such scalars to the enhanced freedom of defining $CP$, 
even beyond the well-known generalized $CP$ symmetries,
which arises in models with several zero-charge scalar fields.
We also show that despite possessing exotic $CP$ quantum numbers,
these scalars do not have to be inert:
they can have $CP$-conserving Yukawa interactions provided the $CP$ acts on fermions by also mixing generations.
This paper focuses on formal aspects --- exposed in a pedagogical manner --- and
includes a brief discussion of possible phenomenological consequences.
\end{abstract}

\maketitle

\section{Introduction}

\subsection{Exploring exotic forms of $CP$-violation}

$CP$-violation was extensively studied experimentally in the past half a century, yet its origin remains enigmatic \cite{book}. 
$CP$-violation may also be present in the leptonic sector, and a vigorous experimental program aims to measure it
\cite{Davies:2011vd,Abe:2011ts,Abe:2013hdq,Messier:2013sfa,Acciarri:2016crz}. 
In the Standard Model (SM), $CP$-violation is introduced by hand in the form of complex quark Yukawa couplings.
In models with extended scalar sectors, it can arise spontaneously, as the result of the Higgs phenomenon,
via $CP$-violating alignment of the Higgs vacuum expectation values (vevs), and
can additionally lead to $CP$-violating Higgs boson exchanges 
\cite{Lee:1973iz,Weinberg:1976hu,Branco:1983tn,2HDM-review,Varzielas:2012nn,Grzadkowski:2013rza,Branco:2015bfb,Fallbacher:2015rea,Emmanuel-Costa:2016vej}. 
Even in the two-Higgs-doublet model (2HDM), a rather conservative extension of the minimal Higgs sector \cite{Lee:1973iz},
the issue of $CP$-violation has many facets \cite{2HDM-review}.
New forms of $CP$-violation were found in the last few years within 2HDM 
\cite{Maniatis:2007de,2HDM-GCP,Maniatis:2009vp,Ferreira:2010bm,Ferreira:2010yh}
and with more than two Higgs doublets \cite{IS-2016,Branco:2015bfb}. 
Understanding how $CP$-violation actually happens may additionally shed some light on the flavor sector hierarchy, 
which is often considered to be intimately intertwined with it,
and on generation of the baryon asymmetry of the universe.

In short, any novel form of $CP$-violation deserves a closer theoretical study as it may tell us something new
and lead to testable predictions.
The present paper is a step towards a deeper understanding of one such unusual form: 
spontaneous violation of the exotic $CP$-symmetry of order four, first proposed in \cite{IS-2016}.
This possibility requires at least three Higgs doublets and was never explored before.
Since it brings up several questions which are not easy to answer, we decided, on the way to its phenomenology, 
to first discuss them in a detailed and pedagogical manner.

\subsection{Order-4 $CP$ symmetry and its consequences}

It is part of our understanding that, besides $CPT$ symmetry, a self-consistent local quantum field theory does not uniquely define any discrete symmetry, such as $C$, $P$, and $T$ transformations. The first general systematic study of this issue was presented long ago by Feinberg and Weinberg \cite{feinberg-weinberg}, but their analysis is restricted to the case of multiplicative phase factors. Generalization to non-abelian discrete groups can be found in \cite{lee-wick}. Further restrictions on phase factors involved in the discrete transformations can be found in \cite{carruthers-1,carruthers-2} and in textbooks, for instance in \citep{book,weinberg-vol1}. 
The freedom of defining the appropriate discrete transformations becomes even larger
in the case of several fields with equal quantum numbers. 
Focusing on the $CP$ transformation acting in the scalar sector with several fields $\phi_i$, $i = 1, \dots, N$, 
one often considers the following generalized $CP$ transformations (GCPs) which we denote by $J_X$:
\be
J_X:\quad  \phi_i(\bx, t) \toCP {\cal CP}\,\phi_i(\bx, t)\, ({\cal CP})^{-1} = X_{ij}\phi_j^{\dagger}(-\bx, t), \quad X_{ij} \in U(N)\,.\label{GCP}
\ee
From now on, we will write $\phi^*$ instead of $\phi^\dagger$ unless the scalars form 
a multiplet such as the electroweak doublets in multi-Higgs-doublet models.
Although the details of the model-building can depend on the matrix $X$, 
the common wisdom is that $J_X$ with any unitary $X$ can play the role of ``the $CP$-transformation'' of the model.
The argument is that all experimentally observable manifestations of $CP$-violation can be related 
to $CP$-violating basis invariant combinations of input parameters and, being basis-invariant, they do not feel the basis change 
induced by $X$ \cite{book}. The same applies to the generalized $T$-transformations, 
which, in the light of the $CPT$ theorem, are also accompanied with the family-mixing matrix.
If the lagrangian and the vacuum of the theory are invariant under any generalized $T$-symmetry, 
there can be no $T$-odd physical observable.

A non-trivial matrix $X$ in (\ref{GCP}) can have peculiar consequences for model-building. 
Note that applying $J_X$ twice leads to a pure family-space transformation:
\be
\phi_i(\bx, t) \to ({\cal CP})^2\phi_i(\bx, t) ({\cal CP})^{-2} = (XX^*)_{ij}\phi_j(\bx, t)\,.\label{GCP2}
\ee
One can bring the matrix $X$ to a block-diagonal form \cite{weinberg-vol1,gcp-standard}, with the blocks
being either $1\times 1$ phases or $2\times 2$ matrices of the following type:
\be
\mmatrix{c_\alpha}{s_\alpha}{-s_\alpha}{c_\alpha}\quad \mbox{as in \cite{gcp-standard},}\quad \mbox{or}\quad
\mmatrix{0}{e^{i\alpha}}{e^{-i\alpha}}{0}\quad \mbox{as in \cite{weinberg-vol1}.}\label{block}
\ee
This is the simplest form of $X$ one can achieve with basis changes in the scalar space $\mathbb{C}^N$.
If $X$ contains at least one $2\times 2$ block with $\alpha \not = \pi$, then 
$J_X^2 = XX^* \not = \mathbb{I}$, so that the $CP$ transformation (\ref{GCP}) 
is not an order-2 transformation.

Until recently, the possibility of having higher-order $CP$ symmetry did not raise much interest.
In all concrete examples considered so far, imposing such a symmetry led to models with other accidental symmetries,
including $CP$ symmetries of order 2. Thus, imposing a higher-order $CP$ was viewed just as a compact way of 
defining a model \cite{2HDM-GCP}, 
not as a path towards {\em new} models that could not be achieved through the usual ``order-2 $CP$ $+$ family symmetry''
combination.
A rare exception is \cite{trautner} where the higher-order $CP$ symmetries were classified 
as distinct opportunities for model building.

The recent work \cite{IS-2016}, developing on an observation made in \cite{abelian}, gave the first concrete example
of a multi-Higgs model in which
the lagrangian is symmetric only under one specific $CP$ symmetry (\ref{GCP}) of order 4 and its powers,
without any accidental symmetry. 
Since this model employs three Higgs doublets, we call it CP4-3HDM and, for completeness, its scalar sector is described in Appendix~\ref{appendix:scalar-CP4-3HDM}.
If  $CP$ is conserved after electroweak symmetry breaking (EWSB),
it leads to a remarkable consequence that the neutral physical scalars can be combined into complex fields 
$\varPhi$ and $\varphi$ that transform under $CP$ in an unusual way:
\be
\varPhi(\bx, t) \toCP i \varPhi(-\bx, t)\,, \quad \varphi(\bx, t) \toCP i \varphi(-\bx, t)\,. \label{varphiCP}
\ee
Notice the absence of complex or hermitian conjugation here.
In order words, the fields $\varPhi$ and $\varphi$ are eigenstates of the $CP$ transformation,
but, unlike the $CP$-even and $CP$-odd fields one usually deals with, they are $CP$-half-odd.

This construction brings in several questions, both fundamental and phenomenological.
Does the absence of conjugation in (\ref{varphiCP}) not render this transformation 
a form of $P$ rather than $CP$ transformation? Does it not lead to any internal inconsistency of the model?
Can such higher-order $CP$ symmetries be detected in the basis-invariant approach?
Do these peculiar scalars lead to any phenomenological signal that cannot be compatible
with any form of the ``standard'' $CP$?

In this work, we will address some of these questions. In Section~\ref{section:scalar-sector} a complete analysis of the scalar sector and its properties under $CP$ is presented. The discussion starts from the ``standard'' $CP$ transformation followed by the concept of  basis dependence. Then the peculiar case of $CP$ transformation without conjugation is explored leading ultimately to the description of the origin and properties of the $CP$-half-odd scalars. In Section~\ref{section:yukawas} the Yukawa interaction is introduced and discussed within the context both of the standard $CP$ symmetry and the order 4 generalized $CP$ case. We show that by extending the idea of generalized $CP$ symmetry to the fermion sector, it is possible to obtain models that obey the symmetry and that contain interactions between $CP$-half-odd scalars and fermions. The Yukawa sectors are then studied before and after electroweak symmetry breaking in 
section \ref{section:yukawas-ewsb}. We summarize our findings and comment on planed work in the conclusions.

\section{The zoo of $CP$ transformations in the scalar sector}
\label{section:scalar-sector}

\subsection{The ``standard'' $CP$ transformation}

The higher-order $CP$-transformations have unconventional consequences,
such as the $CP$-half-odd scalars introduced in \cite{IS-2016}.
In order to accompany the reader through these subtleties,
we begin with a pedagogical introduction to various unusual facets of discrete transformations
acting on scalar fields.
Most of the material in the first half of this section is not new and can be found in textbooks
such as \cite{book} and section 2 of \cite{weinberg-vol1}. 
It is presented here because it will be useful in order to clarify the origin and self-consistency of the $CP$-half-odd scalars,
which will appear towards the end of the section.
In this section, we will deal with purely scalar sector;
interaction of such scalars with fermions will be dealt with in the rest of the paper.

Let us first recap the action of the ``standard'' $C$ and $P$ transformations on scalars. 
Consider a single complex scalar field $\phi(\bx,t)$. After quantization, it is written in terms of creation and annihilation operators that satisfy the standard commutation relations, and it reads
\be
\phi(\bx,t) = \int \tilde{dp} \left[a(\bp) e^{-ipx} + b^\dagger(\bp)e^{ipx}\right]\,,\label{phi}
\ee
where $px \equiv E t - \bp \bx$ and $\tilde{dp} \equiv d^3 p/[2E(2\pi)^3]$
(bold vectors denote 3D momenta or coordinates).
The standard assignment is that the one-particle states 
$a^\dagger(\bp)|0\rangle$ and $b^\dagger(\bp)|0\rangle$ correspond 
to a particle and its antiparticle.\footnote{We stress we need to {\em assign}
what is the antiparticle state for a given particle.
In the case of a single gauge-interacting scalar field, 
we have no other choice but to assign $b^\dagger(\bp)|0\rangle$ as the antiparticle of $a^\dagger(\bp)|0\rangle$.
If we work with two scalars fields with identical quantum numbers,
the freedom to pick up the antiparticle becomes larger, see section~\ref{section-matter-of-choice}
and discussion after Eq.~(\ref{phi1phi2CP3}).}
As a natural consequence of this convention,
one usually postulates that the $C$-transformation acts on operators by exchanging $a$ and $b$,
whereas the $P$ transformation changes the sign of the momentum:
\bea
&&a(\bp) \toC {\cal C}\,a(\bp)\,{\cal C}^{-1} = b(\bp)\,,\quad b(\bp) \toC {\cal C}\,b(\bp)\,{\cal C}^{-1} = a(\bp)\,,\label{standardC}\\
&&a(\bp) \toP {\cal P}\,a(\bp)\,{\cal P}^{-1} = a(-\bp)\,,\quad b(\bp) \toP {\cal P}\,b(\bp)\,{\cal P}^{-1} = b(-\bp)\,.\label{standardP}
\eea
Clearly, both transformations, as well as their product $CP$, are of order 2.
One then immediately sees that, in terms of the original field,
\be
\phi(\bx,t) \toC \phi^*(\bx,t)\,,\quad \phi(\bx,t) \toP \phi(-\bx,t)\,.\label{standard-phi}
\ee
If the lagrangian is invariant under both transformations, they represent the symmetries of the model.
Note also that if it is invariant under the global symmetry group $U(1)$ that rephases $\phi$ to $e^{i\alpha}\phi$
and acts trivially on other fields, then the symmetry (\ref{standard-phi}) arises naturally 
as the non-trivial automorphism of this $U(1)$ symmetry group.

Let us now make a side remark. 
Although one can study how fields transform under $C$ and $P$ transformations separately,
most phenomenologically relevant models, including the SM as well as bSM models with extended scalar sectors,
are chiral, and therefore, they already violate $C$ and $P$ separately through the gauge interactions.
Discussing how scalar fields transform under these separate transformations adds little insight,
but their properties under the combined transformation $CP$ are much more relevant. 
$CP$ conservation or violation does not usually follow from the gauge structure of the models,
and the origin of the small $CP$ violation observed in experiment is puzzling.
So, from now on we will be studying how fields transform under the combined transformation $CP$,
without splitting it into $C$ and $P$, which is in any case not uniquely determined.
Moreover, starting from the next subsection, we will suppress the arguments
both of the fields and of the operators. It is always assumed that, for fields, $\bx \to - \bx$ 
and, for operators, $\bp \to -\bp$ under the $CP$ transformation.

Writing $\phi(\bx,t)$ via two real fields, $\phi = (h_1 + i h_2)/\sqrt{2}$,
we see that $h_1$ is $CP$-even and $h_2$ is $CP$-odd. Expressing the two real fields via operators,
\be
h_{1,2}(\bx,t) = \int \tilde{dp} \left[a_{1,2}(\bp) e^{-ipx} + a_{1,2}^\dagger(\bp)e^{ipx}\right]\,,\label{h1h2}
\ee
we identify
\be
a_1 = {a+b \over \sqrt{2}}\,,\quad a_2 = {a-b \over \sqrt{2}i}\,,
\ee
and see that 
\be
a_1(\bp) \toCP a_1(-\bp)\,,\quad a_2(\bp) \toCP - a_2(-\bp)\,.\label{a1a2CP}
\ee
If the original lagrangian in terms of $\phi$ was invariant under $U(1)$, 
then, in the space of $h_1$ and $h_2$, it is invariant under $O(2) \simeq SO(2)\rtimes \Z_2$,
where the original $U(1)$ symmetry is mapped onto $SO(2)$, while the extra $\Z_2$ transformation
is given by the sign flip of $h_2$.

Certainly, if $\phi$ is charged under gauge interactions, so that the single-particle states $a^\dagger |0\rangle$ 
and $b^\dagger |0\rangle$ differ by their conserved charges,
then it makes little sense to switch to the two real fields $h_1$ and $h_2$.
The conserved charge operator, together with the hamiltonian, fixes the most convenient basis to work in.
However, in absence of any gauge interactions, working with operators $a_1$, $a_2$ or with $a$, $b$
becomes just a matter of convention.

Finally, we can revert the flow of the arguments. Suppose we have a model with two real mass-degenerate fields $h_1$ and $h_2$,
one of them being $CP$-even, the other being $CP$-odd.
This assignment can, for example, arise as a result of rearrangement of scalar degrees of freedom after spontaneous symmetry breaking, 
or it can be imposed by hand on operators (\ref{a1a2CP}) in toy models.
Then we are allowed to rewrite it in terms of a single complex field $\phi$ transforming under $CP$ as $\phi(\bx,t) \toCP \phi^*(-\bx,t)$.
This remark sounds trivial, but we will see below how similar arguments lead to less familiar conclusions.

\subsection{Basis dependence}\label{section-basis-dependence}

Suppose $\phi\toCP \phi^*$ as before. Let us make a basis change and define a new scalar field as
$\phi' = i \phi$. Then 
\be
\phi'\toCP i \phi^* = - (i \phi)^* = - (\phi')^*\,.\label{varCP}
\ee
We stress the all-important property that $CP$ transformation is unitary, not antiunitary, 
and therefore $i$ stays intact in the first transition.
The rephased complex field transforms in a different
way under {\em the same} $CP$ transformation. 
Now, writing $\phi' = (h_1' + i h_2')/\sqrt{2}$, we see that it is the real part, $h_1'$, 
which is $CP$-odd and the imaginary part, $h_2'$, which is $CP$-even.
We come to the well-known conclusion that the exact form of the $CP$ transformation law is basis-dependent.

It may happen that the model possesses {\em another} $CP$ symmetry, $CP'$,
under which $\phi$ transforms with an extra minus sign, while the rephased field $\phi'$ transforms 
in the ``standard'' way.
Both $CP$ and $CP'$ can play the role of ``the'' $CP$-symmetry of the model.
Then, even in a fixed basis, there is no unique assignment for which degree of freedom is $CP$-even and which is $CP$-odd.
One must specify the particular $CP$ transformation one wishes to test.
Notice also that the product of $CP$ and $CP'$, which acts by just flipping the sign $\phi \to -\phi$,
is also a symmetry. This is a straightforward feature of models incorporating more than one $CP$ symmetry.
A famous example of this situation is found in the Inert doublet model (IDM), a version of two-Higgs-doublet model (2HDM)
with an exact $\Z_2$ symmetry which flips the sign of the second, inert, Higgs doublet \cite{IDM-1,IDM-2,IDM-3,IDM-4}. 
This inert doublet gives rise to heavy neutral scalars $H \propto \Re\, \phi_2^0$ and $A \propto \Im\, \phi_2^0$, 
which are known to be of opposite $CP$ parity but we cannot uniquely assign which is which. 

In Eq.~(\ref{varCP}), we used a very specific basis change to illustrate that the ``standard'' $CP$ transformation rule
is basis-dependent. A generic basis change leads to 
\be
\phi' \toCP e^{i\alpha} (\phi')^*\,.\label{varCP2}
\ee 
We stress once again that this is {\em not} a new $CP$ transformation;
it is the same transformation seen in a different basis.

\subsection{$CP$-transformation without conjugation}\label{section-without-conjugation}

Consider now a model with two real, mass-degenerate scalars $h_1$ and $h_2$, both of which are $CP$-even: 
$h_1 \toCP h_1$ and $h_2 \toCP h_2$. 
By saying that, we assume that these scalars represent only a part of the full theory,
and it is the full theory that prevents any other assignment for the $CP$-transformation.
We are then allowed to combine them 
into a single complex field $\phi \equiv (h_1 + i h_2)/\sqrt{2}$.
By definition, it is $CP$-even, $\phi \toCP \phi$,
and no conjugation is involved under the action of $CP$.
The possibility of having $CP$-even complex scalar field is not new, see, for example, Eq.~(23.41) in \cite{book}. 

This simple math necessitates the following interpretation: the field $\phi$ is self-conjugate under $CP$ and, therefore, under $C$, up to a possible phase factor.
So is $\phi^*$; the two are not related through the $CP$-transformation we started with.
In terms of creation and annihilation operators, by using (\ref{h1h2}), we build $\phi$ defined as in (\ref{phi}) 
with the following operators:
\be
a = {a_1+ i a_2 \over \sqrt{2}}\,,\quad b = {a_1 -i a_2 \over \sqrt{2}}\,.
\ee
Keeping in mind that these operators are mapped under $CP$ as 
$a(\bp) \toCP a(-\bp)$, $b(\bp) \toCP b(-\bp)$, 
we see again that $a^\dagger|0\rangle$ and $b^\dagger|0\rangle$
correspond to two {\em different} particles, not a particle and its antiparticle, 
as is usually implied when writing (\ref{phi}).
Now, since the (unbroken) gauge symmetry assigns opposite gauge charges to particles
and antiparticles, we conclude that the $CP$-even complex scalar $\phi$ 
cannot possess any conserved non-zero gauge quantum number 
that would tell particle from antiparticle.
Only when this condition is fulfilled, an ambiguity exists in defining how 
the charge conjugation acts in this sector.

Notice that the standard canonical quantization procedure,
as well as the computation of the hamiltonian and momentum density
in terms of operators $a$ and $b$, remain exactly as they are in the case 
of the usual complex field. This computation is based only on the algebraic manipulation
of operators but does not rely on any interpretation relating the two.

\subsection{Conjugating or not under $CP$ is a matter of basis choice}\label{section-matter-of-choice}

The construction made above may seem artificial and one may suspect
that models based on respectfully looking, generalized $CP$ transformations of the form (\ref{GCP}) 
never involve such peculiarities. We will now show that they do.
Just as in section~\ref{section-basis-dependence}, where we showed that the presence of the minus sign in the definition
of the $CP$ transformation is a matter of basis choice, we will now show that the conjugation involved in the usual $CP$ transformation can be ``undone'' in certain situations.

Consider two complex scalar fields, $\phi_1$ and $\phi_2$, with definite masses transforming under $CP$ as
\be
\phi_1 \toCP \phi_2^*\,,\quad \phi_2 \toCP \phi_1^*\,.\label{phi1phi2CP3}
\ee
If this transformation has a chance to represent a symmetry of the model, 
$\phi_1$ and $\phi_2$ must be mass-degenerate and have identical gauge quantum numbers.
In short, they must form a multiplet of complex scalar fields.
Expressing them via operators (\ref{phi})
we obtain that, under the $CP$-transformation, $a_1(\bp) \leftrightarrow b_2(-\bp)$ and $a_2(\bp) \leftrightarrow b_1(-\bp)$.
Thus, we already encounter a situation similar to the previous subsection: 
the one-particle state $a_i^\dagger|0\rangle$ and $b_i^\dagger|0\rangle$, for the same $i$,
are not particle and antiparticle of each other,
despite these two operators residing inside the same field.
These are, instead, two distinct particles, albeit with the same mass and 
opposite gauge charges.

Next, perform a $\pi/4$-rotation in the space of complex fields:
\be
\doublet{\eta}{\xi} = {1 \over \sqrt{2}}\mmatrix{1}{1}{-1}{1}\doublet{\phi_1}{\phi_2}\label{12-to-eta-xi}
\ee
and observe that, upon $CP$, the new fields transform as $\eta \toCP \eta^*$,
$\xi \toCP -\xi^*$, in a very conventional way. This is also manifest at the level of creation and annihilation
operators; for example, $a_1+a_2$ residing inside the field $\eta$ indeed turns into $b_1+b_2$ upon $CP$ transformations,
just as expected for a usual complex field.

Now, among the four real degrees of freedom, there are two $CP$-even, $\Re\eta$ and $\Im\xi$,
and two $CP$-odd, $\Re\xi$ and $\Im\eta$.
Since they correspond to mass-degenerate fields, one can recombine fields with the same
$CP$ parity into the new complex fields $\Phi = \Re\eta - i\,\Im\xi$ and 
$\tilde \Phi = \Re\xi - i\, \Im\eta$.
In this way, the new fields become, as in the previous subsection, 
the $CP$-even and $CP$-odd complex fields: 
\be
\Phi \toCP \Phi\,,\quad \tilde \Phi \toCP - \tilde \Phi\,.\label{Phi-Phi-tilde}
\ee
Linking $\eta$ and $\xi$ to the original complex fields, 
we can express the passage from $\phi_1$ and $\phi_2$ to $\Phi$ and $\tilde\Phi$
as the following transformation $R$:
\be
R: \quad \doublet{\Phi}{\tilde\Phi} = {1 \over \sqrt{2}}\mmatrix{1}{1}{-1}{1}\doublet{\phi_1}{\phi_2^*}\,.
\label{12-to-Phi-Phitilde}
\ee
This is the basis change which ``undoes'' the conjugation under $CP$.

Let us see what happens from the algebraic point of view.
In Eq.~(\ref{12-to-eta-xi}), we pass from the $(\phi_1,\phi_2)$ to the $(\eta,\xi)$-description of the same space
$\mathbb{C}^2$ in a way that preserves its complex structure (holomorphic map). 
This is the usual basis change belonging to $U(2)$.
In Eq.~(\ref{12-to-Phi-Phitilde}), when passing from $(\phi_1,\phi_2)$ to $(\Phi,\tilde\Phi)$, 
we also map the same $\mathbb{C}^2$ onto itself, but via a non-holomorphic map 
(notice $\phi_2^*$ instead of $\phi_2$).
The transformation $R$ cannot be represented by any $U(2)$ transformation.
However it does conserve the norm of the vector and it belongs to the group $O(4)$ of rotations
in $\mathbb{R}^4$ spanned by $(\Re \eta, \Im \eta, \Re \xi, \Im \xi)$ at each space-time point.
Denoting the corresponding operators for the fields $\Phi$ and $\tilde\Phi$
as $a, b$ and $\tilde a, \tilde b$, respectively,
we establish the following relation:
\be
\left(\begin{array}{c} a \\ \tilde a \\ b\\ \tilde b\end{array}\right) = 
{1 \over \sqrt{2}}
\left(\begin{array}{rrrr} 
1 & 0 & 0 & 1 \\ 
-1 & 0 & 0 & 1 \\
0 & 1 & 1 & 0 \\
0 & 1 & -1 & 0
\end{array}\right) \left(\begin{array}{c} a_1 \\ a_2 \\ b_1\\ b_2\end{array}\right) \,.\label{Rab}
\ee
Promoting the basis change group from $U(2)$ to $O(4)$ is allowed only 
if there are no gauge quantum numbers that distinguish $\phi_i$ from $\phi_i^*$.
In this case, we are free to {\em define} what we call particle and antiparticle.
In (\ref{phi1phi2CP3}), we assumed $\phi_2$ to be the fundamental field, and labeled its conjugate as $\phi_2^*$.
But since they do not differ in their quantum numbers, we could have reversed the notation from the very start.
In this case, the creation operators inside $\phi_1^*$ and $\phi_2^*$ would correspond to a particle-antiparticle pair,
creation operators inside $\phi_1$ and $\phi_2$ --- to {\em another} particle-antiparticle pair.
The interpretations of Eq.~(\ref{12-to-eta-xi}) and Eq.~(\ref{12-to-Phi-Phitilde}) would also be reversed:
the latter would be the normal basis change, which the former would be a non-holomorphic transformation.
Let us stress once again that, when doing these manipulations, {\em we never redefine} the $CP$ transformation itself.
We only make the basis changes, keeping the same $CP$ transformation all the way.\footnote{Let us stress that although $R$ mixes $\phi$'s and their conjugates,
it is certainly {\em not} a Bogolyubov transformation, as it does not mix
the creation and annihilation operators. The extended symmetry group allows us
to mix $a$'s not only with other $a$'s but also with $b$'s, but never with $a^\dagger$ or $b^\dagger$.
The transformation $R$ acting on $a$'s and $b$'s as in (\ref{Rab}) is unitary,
and as a result, the canonical commutation relations and the normal ordering are always preserved.}

The lesson from this discussion is the following: in the case of two complex scalar fields possessing no charges,
the distinction between particles and antiparticles is blurred to such extent that the space
of fields acquires a larger intrinsic basis change freedom.
All mutually orthogonal one-particle states can still be grouped into pairs of particles and antiparticles,
but we have a certain freedom of defining them.
As the result, the $C$-transformation loses its importance as a  transformation that maps particles
to antiparticles, and in certain basis $CP$ can look just like $P$-transformation.
Note importantly that it is not about redefining the $CP$ transformation, it is about the basis change freedom:
{\em the same} $CP$-transformation appears as the usual one in one basis and a $P$-resembling transformation in another.

\subsection{Order-4 $CP$-transformation and $CP$-half-odd states}\label{section-CP4}

The ambiguity of choosing degrees of freedom in models with two complex scalar fields not participating in gauge interactions
allows one to implement even more exotic features.
Consider, instead of (\ref{phi1phi2CP3}), the following $CP$ transformation:
\be
J: \quad \phi_1 \toCP i\phi_2^*\,,\quad \phi_2 \toCP -i\phi_1^*\,.\label{order4-example}
\ee
This transformation closely matches the $CP$-transformation used in the CP4-3HDM, see (\ref{Jb}).
The conjugate fields transform, naturally, as
$\phi_1^* \toCP -i\phi_2$, $\phi_2^* \toCP i\phi_1$. 
The transformation defined in (\ref{order4-example}) is of order 4, as $J^2 \not = \mathbb{I}$, $J^4= \mathbb{I}$.
The $CP$ transformation acts on the operators $a_i$, $b_i$, in the following way:
\be
a_1 \toCP i b_2\,,\ b_2 \toCP i a_1\,, \quad a_2 \toCP -i b_1\,,\ b_1 \toCP -i a_2\,.
\ee
Once again, in order for the transformation law to represent a symmetry of at least
the free theory, the two fields $\phi_1$ and $\phi_2$ must possess equal quantum numbers and be mass-degenerate.

Repeating the previous analysis, we regroup the four real degrees of freedom
in the following way:
\be
\doublet{\varPhi}{\varphi^*} = {1 \over \sqrt{2}}\mmatrix{1}{1}{-1}{1}\doublet{\phi_1}{\phi_2^*}\,.
\label{12-to-varphis}
\ee
Notice that, in contrast to (\ref{12-to-Phi-Phitilde}), we changed the definition of the second field $\varphi$.
We could have done it in the previous example as well, at the expense of a slightly longer discussion.
The new fields $\varPhi$ and $\varphi$ transform under $CP$ in a remarkable way:
\be
\varPhi \toCP i \varPhi\,, \quad \varphi \toCP i \varphi\,,
\ee
Notice again the disappearance of the conjugation.
At the level of operators, one has
\be
a_\varPhi = {a_1 + b_2 \over \sqrt{2}} \toCP i a_\varPhi\,,\quad 
b_\varPhi^\dagger = {a_2^\dagger + b_1^\dagger \over \sqrt{2}} \toCP i b_\varPhi^\dagger\,,
\ee
and similarly for $a_\varphi = (a_2-b_1)/\sqrt{2}$ and $b_\varphi^\dagger = (b_2^\dagger - a_1^\dagger)/\sqrt{2}$.

The new fields are $CP$-eigenstates.
Therefore, one can associate with them a quantum number $q$, which is defined modulo 4 
and which generalizes the notion of $CP$ parity.
We have $q_\varPhi = q_\varphi = +1$, $q_{\varPhi^*} = q_{\varphi^*} = -1$.
In terms of single-particle states, $b^\dagger_\varPhi|0\rangle$ and $b^\dagger_\varphi|0\rangle$
have $q = +1$, while $a^\dagger_\varPhi|0\rangle$ and $a^\dagger_\varphi|0\rangle$
have $q = -1$.
If the model contains a $CP$-odd field, it must be associated with $q = 2$ (the sign is irrelevant).
Although $\varPhi$ and $\varphi$ are eigenstates of the $CP$-transformation,
they are neither $CP$-even nor $CP$-odd, but rather $CP$-half-odd.
This is essentially what was found in the CP4-3HDM \cite{IS-2016}.
The origin of such states is (again) the extra freedom of basis change 
that one gets for two complex fields with identical masses and zero gauge couplings.

\subsection{$CP$-eigenstates are not compatible with conserved global charges}

The constructions presented in the preceding three subsections
demand the following observation.\footnote{We thank the referee for raising this issue.}
Consider first the situation described in section~\ref{section-without-conjugation}
with two mass-degenerate real fields $h_1$ and $h_2$ in the {\em free field limit}. 
The lagrangian acquires the $O(2)$ symmetry group that via Noether's theorem 
leads to the conserved global charge operator
\be
Q = \int \tilde{dp}\, a_i^\dagger(\bp) T_{ij} a_j (\bp) = \int \tilde{dp}\, [a^\dagger(\bp)a(\bp) - b^\dagger(\bp) b(\bp)]\,,
\label{charge-Q}
\ee
where $T_{ij} = i \epsilon_{ij}$ is the $SO(2)$ generator.
The one-particle states $a^\dagger|0\rangle$ and $b^\dagger|0\rangle$ 
are eigenstates of this operator with charges $\pm 1$, respectively.

Had we used the usual assignment for $CP$, even accompanied with the phase factors, 
we would see $a^\dagger a \leftrightarrow b^\dagger b$ under $CP$, and as the result
\be
(CP) Q (CP)^{-1} = - Q\,.\label{usual-charge}
\ee
This is compatible with the notion that $CP$ turns particles with given (global) charges
into antiparticles with opposite charges.
However, if we stick to the $CP$-transformation suggested in section~\ref{section-without-conjugation},
the one under which both $h_1$ and $h_2$ are even and, equivalently, $\phi \toCP \phi$
without conjugation, we get $(CP) Q (CP)^{-1} = Q$.
In simple words, the so-defined $CP$ does {\em not} flip the sign of the global charge. 
One can legitimately question the validity of this definition as $CP$.

A similar situation takes place for two mass-degenerate complex fields $\phi_1$ and $\phi_2$
as described in sections~\ref{section-matter-of-choice} and \ref{section-CP4}.
Again, in the free-field limit, one has the global symmetry group $SO(4)$ of rotations among
the four real degrees of freedom that leads to six charge operators $Q_a$, with expressions
similar to the first expression in (\ref{charge-Q}) but with six $SO(4)$ generators inserted.
The Cartan subalgebra of $so(4)$ is two-dimensional, and we need to pick up
two commuting charges out of six to classify one particle states.
The exact choice of these two charge operators depends 
on how we combine the real degrees of freedom to build one-particle states. 
For example, in terms of the operators inside the complex fields, $a_1$, $a_2$, $b_1$, $b_2$,
we can define two charges
\be
Q_{1,2} = \int \tilde{dp}\left[a_1^\dagger a_1 - b_2^\dagger b_2 
\pm \left(a_2^\dagger a_2 - b_1^\dagger b_1\right)\right]\,.
\label{charges-Q12}
\ee
The physically valid definition of $CP$-transformation is required only to invert the signs
of those charge operators which are used to classify the states.
The charge operators $Q_1$ and $Q_2$ in (\ref{charges-Q12}) are chosen so that
they indeed change signs not only under the ``standard'' $CP$-transformation 
$a_i \toCP b_i$ but also the GCP used in section~\ref{section-matter-of-choice} 
with $a_1 \toCP b_2$ and $a_2 \toCP b_1$. Thus, in the free-field theory, 
this definition of GCP is compatible
with particle and antiparticles having opposite conserved charges.

The exotic complex fields $\varPhi$ and $\tilde\varPhi$ are also eigenstates
of two commuting charge operators; they are different from those of Eq.~(\ref{charges-Q12})
but acquire the same form if written in terms of $a, b, \tilde a, \tilde b$ defined in Eq.~(\ref{Rab}).
The problem is, however, that these charge operators {\em do not change signs} under the same GCP.
The same conclusion holds for the order-4 GCP of section~\ref{section-CP4}.
This is the general consequence of $\varPhi$ and $\tilde\varPhi$ being $CP$-eigenstates:
since $a$'s and $b$'s are not swapped under GCP,
all combinations such as $\tilde a^\dagger \tilde a$ stay invariant.

The only way out is to demand that there be {\em no conserved global charge operators} 
whose eigenstates could be $CP$-eigenstates. In our original model CP4 3HDM,
as well as in the pedagogical examples considered above, 
we always require that in the full theory there be no accidental continuous symmetries and, therefore, no conserved charges.
In particular, within CP4 3HDM, it is guaranteed by the self-interaction terms.
In this situation, no problem raised in this subsection arises.
But in the free field theory limit, when the conserved global charges appear,
the non-conjugating transformation loses its status as a $CP$-transformation.
This discontinuous transition is fine, since the free theory is indeed qualitatively different from
the interacting one, at least in what concerns the structure in the Hilbert space of states.

In short, a non-conjugating $CP$-transformation is incompatible with scalars that possess conserved charges,
either gauge or global. The exotic $CP$-half-odd scalars are possible only in theories without continuous symmetries.

\subsection{Further remarks}

Before ending this section we present a few additional comments that sum up the situation for scalar fields.

First, notice that for the construction we have presented above, it is essential to have {\em two} complex fields: in order for the transformation (\ref{order4-example}) to work, $a_i$ and $b^\dagger_i$ must transform in the same way.
Trying to impose a $CP$ transformation similar to (\ref{order4-example}) for a single field,
$a \toCP ib,\, b\toCP ia$, would lead to
\be
\phi\toCP  \phi_{CP} = i \int \tilde{dp}\, [b(\bp) e^{-ipx'} - a^\dagger(\bp) e^{ipx'}]\,,\label{pathological}
\ee
where $x' = (t, -\bx)$. The field $\phi_{CP}$ cannot be written as a linear combination of 
$\phi$ and $\phi^*$.
As a result, the hamiltonian density ${\cal H}(\phi, \phi^*)$ transforms into
${\cal H}(\phi_{CP}, \phi^*_{CP}) \not = {\cal H}(\phi, \phi^*)$.
Even if one tries to construct a hamiltonian containing both $\phi$ and $\phi_{CP}$ as in (\ref{pathological})
as well as their conjugates, one would find that $[\phi(x), \phi_{CP}(y)]$ does not vanish at space-like separation
$(x-y)^2 < 0$.
Thus, this symmetry cannot be conserved in any local causal quantum field theory.

It is nice to notice that this kind of trouble was observed by Carruthers back in 1967 \cite{carruthers-1}
as an unavoidable consequence of introducing self-conjugate half-integer-isospin multiplets of boson fields.
For example, for the isospin $T = 1/2$ one can define the boson field with $T_3 = +1/2$ 
with operators $a$ and $b^\dagger$ as usual. If the conjugate state of this boson
belongs to the same multiplet, with $T_3 = -1/2$, then the isospin conservation dictates
that there appears an extra minus sign between the exchanged operators, just as in (\ref{pathological}).
As a result, the commutator of these two boson fields 
does not vanish at space-like distances, rendering the theory non-local.
Thus, if one has a theory with bosons sitting in an isospin doublet, 
then their conjugates must form another multiplet, as it happens for kaons even disregarding their charges.

Second, we stressed above that, in order for the non-holomorphic basis change to work,
the scalars must have zero charges, so that neither gauge coupling nor a conserved global quantum number
could distinguish a particle from an antiparticle.
In the context of multi-Higgs-doublet models, and in particular the CP4-3HDM,
this condition is naturally satisfied for the neutral component of the scalar fields
that do not acquire the vacuum expectation values. Indeed, after EWSB, we are left only with the electromagnetic gauge group,
to which the neutral Higgses do not couple. In short, there is no need to introduce
$CP$-half-odd scalars by hand; they naturally arise in certain multi-doublet models.

Third, please note that there is certain resemblance between our treatment and the ``Majorana formalism''
developed in \cite{pilaftsis-1,pilaftsis-2,pilaftsis-3,maas-pedro} for 2HDM, and to the more general approach to $CP$ symmetries
in extended scalar sectors presented in \cite{lindner}.
The key similarity to both works is to combine $\phi$'s and $\phi^*$'s into a single
multiplet $\Phi$. The effect of $CP$ --- standard or generalized --- on $\Phi$
is just a transformation of $\Phi$. This is so because $\Phi^*$ is not an independent field
anymore but can be expressed as a linear map of $\Phi$. It is this property, or to be precise,
its specific realization in 2HDM, that was called in \cite{pilaftsis-1} the ``Majorana property''
for scalars. 

Also, if one neglects the $U(1)_Y$ part of the electroweak gauge
group, then the doublets $\phi_i$ and $\tilde\phi_i = i \sigma_2 \phi^*_i$ transform 
in the same way under $SU(2)_L$, and can indeed be arbitrarily mixed \cite{pilaftsis-1,pilaftsis-2,pilaftsis-3,maas-pedro}
without spoiling the kinetic term and gauge interactions.
This is reminiscent of our observation that when the fields $\phi$ possess no conserved charges, 
one gets an enhanced transformation freedom.
The difference is that the electroweak symmetry breaking, at least in its perturbative formulation after gauge fixing,
breaks the Majorana construction of \cite{pilaftsis-1,pilaftsis-2,pilaftsis-3,maas-pedro},
while in our case, and specifically in the CP4-3HDM, it survives and affects the $CP$-properties
of the physical neutral scalars.

Fourth, a well-known computation shows that the action of $T$ transformation
squared on any single-particle state amounts to the factor $(-1)^{2j}$, where $j$ is its spin \cite{weinberg-vol1}.
This leads to the famous theorem by Kramers \cite{kramers} that in any system with an odd 
number of fermions described by a $T$-invariant hamiltonian the energy eigenstates must be double degenerate.
This feature is known as Kramers degeneracy.
This relation, however, gets modified in the presence of mass-degenerate multiplets \cite{weinberg-vol1}.
When $T$ acts on a single-particle state it can, in addition to flipping momentum and helicity,
also map it to another single-particle state.
The action of $T^2$ is then given by $(-1)^{2j} e^{2i\alpha}$, and for $\alpha=\pi/2$
it leads to a sign factor {\em opposite} to the standard result.
This possibility requires a two-fold degeneracy of mass eigenstates beyond Kramers doubling.

Fifth and final: recall that the one-particle state $b_\varPhi^\dagger|0\rangle$ associated with the CP-half-odd
complex field $\varPhi$ is its own antiparticle, up to the extra $i$ factor accompanying the action of $CP$.
The one-particle states $a^\dagger|0\rangle$ arising from $\varPhi^*$ 
is a {\em different} one-particle state and is also its own antiparticle.
The presence of the extra $i$ factor leads to the remarkable prediction 
that a pair of such bosons possesses the ``wrong'' $CP$-parity.
In their center of motion frame
\be
({\cal CP})a_\varPhi^\dagger(\bp)a_\varPhi^\dagger(-\bp)({\cal CP})^{-1} =
- a_\varPhi^\dagger(\bp)a_\varPhi^\dagger(-\bp)\,.
\ee
Since the operators $a_\varPhi$ satisfy the usual commutation relations,
such a pair must sit in an even partial wave state. However its intrinsic $CP$-parity is negative.
Thus, we obtain a peculiar situation of a $CP$-odd pair of two identical bosons.
This is something that is usually considered impossible for bosons and 
that was encountered so far only for majorana fermions.

We note in passing\footnote{We are grateful to Leonardo Pedro who brought our attention to that work.} 
that in a more mathematically refined formalism the quantum fields can be defined according
to their transformations under Pin rather than Spin groups \cite{pin}. 
In this formalism the notions of discrete transformations must also be adjusted,
one can discuss fermion states with have ``parity'' $\pm i$ rather than $\pm 1$.
It is not clear to us whether there is a deeper connection between the two phenomena.

\section{Coupling $CP$-half-odd scalars to fermions}
\label{section:yukawas}

The scalar coupling with (charged) fermions is described via Yukawa interactions. 
In this section, we want to investigate whether this can be done for $CP$-half-odd scalars
in the $CP$-conserving fashion.
The textbook classification of single-fermion (pseudo)scalar bilinears being $CP$-even and/or $CP$-odd might lead one to suspect that it is impossible to couple them to $CP$-half-odd scalars and thus that such scalars must be genuinely inert.
We will show, however, that this is true only if $CP$ acts
on fermions in the traditional --- fermion-family-blind --- way. If, instead, the $CP$-transformation
mixes fermion generations, just like it mixes scalars, then $CP$ conserving Yukawa interactions are allowed.

\subsection{Yukawa sectors with an order-2 $CP$-symmetry}

To keep the exposition pedagogical, let us start with the most general Yukawa sector 
with $n_f$ fermion fields $\psi_i$ coupled to a single neutral complex scalar field $\phi$: 
\be
{\cal L} = \bar \psi_i (A_{ij} + B_{ij}\gamma_5)\psi_j\, \phi + 
\bar \psi_i [(A^\dagger)_{ij} - (B^\dagger)_{ij}\gamma_5]\psi_j\, \phi^*\, ,\label{L1}
\ee
with arbitrary complex $n_f \times n_f$ matrices $A$ and $B$. 
We assume for the moment that the $CP$ transformation acts on fermions in the standard way, up to an overall phase which is the same for all generations,
so that bilinears get transformed as
\be
\bar \psi_i \psi_j \toCP \bar \psi_j \psi_i\,,\quad 
\bar \psi_i \gamma_5 \psi_j \toCP - \bar \psi_j \gamma_5\psi_i\,. \label{cp-ferm-1}
\ee
With this convention, Eq.~(\ref{L1}) transforms into
\be
{\cal L} = \bar \psi_i (A^T_{ij} - B^T_{ij}\gamma_5)\psi_j\, \phi_{CP} + \bar \psi_i [(A^*)_{ij} + (B^*)_{ij}\gamma_5]\psi_j\, \phi^*_{CP}\,.\label{L1CP}
\ee
We want this Yukawa sector to be $CP$-conserving, 
which requires that we specify how the $CP$-transformed scalar $\phi_{CP}$ is related with $\phi$.
The conventional transformation law $\phi \toCP \phi_{CP} = \phi^*$ immediately forces
both matrices, $A$ and $B$, to be real. 
Moreover, writing as usual $\phi = (h_1 + i h_2)/\sqrt{2}$
with the $CP$-even real field $h_1$ and $CP$-odd $h_2$ and decomposing
Yukawa matrices into symmetric and antisymmetric parts, $A_{s,a} = (A\pm A^T)/\sqrt{2}$, $B_{s,a} = (B\pm B^T)/\sqrt{2}$,
we get
\be
{\cal L} = \bar \psi_i (A_s+B_a \gamma_5)_{ij}\psi_j\, h_1 + i \bar \psi_i (A_a+B_s \gamma_5)_{ij}\psi_j\, h_2.\label{ABas}
\ee
In particular, for a single fermion generation, we recover the traditional expression
\be
{\cal L} = \sqrt{2} A\, \bar \psi \psi \cdot h_1 + i \sqrt{2} B\, \bar \psi \gamma_5\psi \cdot h_2\label{L2CP}
\ee
with real $A$ and $B$. 

Allowing for extra rephasing upon $CP$-transformation of $\phi$, such as in Eq.~(\ref{varCP2}), 
produces no effect on this construction, since it can be removed with a basis change
accompanied by the overall phase change of $A$ and $B$.
The conclusion is that the Yukawa sector has a $CP$-symmetry
if all entries in the the matrices $A$ and $B$ have the same phase.
In particular, even in the single-generation case (\ref{L2CP}),
if couplings $A$ and $B$ fail to satisfy $\Im AB^* = 0$,
then the same scalar degree of freedom will couple both to $\bar\psi\psi$ and to $\bar\psi\gamma_5\psi$.
This makes the Yukawa sector $CP$-violating even with one fermion generation.

In Section~\ref{section-matter-of-choice} we argued that a model with two mass-degenerate
scalar fields with zero gauge quantum numbers enjoys a larger group of basis changes,
which allowed us to recast the conventional $CP$-transformation in the form (\ref{Phi-Phi-tilde}).
Let us now see how this extended basis change affects the Yukawa sector.

We start with the general Yukawa sector as in (\ref{L1}) and duplicate it, with matrices $A_1$ and $B_1$
corresponding to $\phi_1$ and matrices $A_2$ and $B_2$ corresponding to $\phi_2$.
Assuming that the $CP$ transformation acts on scalars as in (\ref{phi1phi2CP3}), we deduce from $CP$-conservation 
that $A_2^* = A_1 \equiv A$ and $B_2^* = B_1 \equiv B$.
Let us focus on the part of the lagrangian that couples scalars to $\bar\psi_i \psi_j$.
Omitting indices, we perform the following regrouping:
\bea
A\phi_1 + A^*\phi_2 + h.c. &=& A\phi_1 + A^*\phi_2 + A^\dagger \phi_1^* + A^T\phi_2^* \nonumber\\
&=& {A + A^T \over \sqrt{2}} \cdot {\phi_1 +\phi_2^* \over \sqrt{2}} - {A - A^T \over \sqrt{2}} \cdot {-\phi_1 +\phi_2^* \over \sqrt{2}} + h.c.\nonumber\\
&=& A_s \Phi - A_a \tilde\Phi + h.c. 
\eea
Here, we used the complex scalar fields $\Phi$ and $\tilde\Phi$ defined in (\ref{12-to-Phi-Phitilde}), 
whose $CP$-parities are given in (\ref{Phi-Phi-tilde}), and introduced symmetric and antisymmetric
parts of $A$: $A_s = (A+A^T)/\sqrt{2}$, $A_a = (A-A^T)/\sqrt{2}$.
Repeating it for matrices $B$, we arrive at the following Yukawa sector in terms of fields $\Phi$ and $\tilde\Phi$:
\be
{\cal L} = \bar\psi_i \left[(A_s + B_a\gamma_5)_{ij}\Phi - (A_a + B_s\gamma_5)_{ij}\tilde\Phi\right]\psi_j + h.c.\label{AsBa}
\ee
The (anti)symmetric parts of coupling matrices exactly match the $CP$ properties of $\Phi$ and $\tilde\Phi$.
Notice also that this expression resembles (\ref{ABas}) with the exception that the $CP$-even and $CP$-odd scalar fields
are now complex and, as a consequence, we do not need to impose any phase condition on matrices $A$ and $B$.
Notice also that we could have constructed (\ref{AsBa}) directly from (\ref{L1}) 
just by using the known $CP$-properties of the new scalar fields.

\subsection{Yukawa sectors with an order-4 $CP$-symmetry}

We now turn to the order-4 $CP$-symmetry 
and try to couple the $CP$-half-odd scalar $\varPhi$ to fermions.
We first assume that the $CP$ transformation acts on fermions in the conventional way.
Then, starting with (\ref{L1}) and using $\varPhi \toCP i \varPhi$, we arrive at the conditions
\be
i A^T = A\,,\quad -i B^T = B\,.\label{noway}
\ee
Applying twice, we get $A = -A$, $B= - B$, which sets both of them to zero.
The only way towards a non-zero coupling of $CP$-half-odd scalar to fermions
is to assume that the $CP$-transformation acts non-trivially on $n_f$ fermion generations:
\be
\psi_i \toCP Y^*_{ij}\cdot i\gamma^0 {\cal C} \bar \psi_j^T\,, \quad Y \in U(n_f)\,,\label{Y}
\ee
where, for definiteness, we selected a specific phase convention.
Note that if we aim at constructing a model that preserves the so-defined $CP$ in all sectors,
we must require the fermions participating in the family mixing to be mass-degenerate.
With these conventions, the fermion bilinears transform as
\be
\bar\psi_i(A_{ij} + B_{ij}\gamma_5)\psi_j \toCP \bar\psi_{j'}Y^*_{jj'}(A_{ij} - B_{ij}\gamma_5)Y_{ii'}\psi_{i'} =
\bar\psi_i [Y^\dagger (A^T - B^T\gamma_5)Y]_{ij}\psi_j\,.
\ee
Therefore, instead of (\ref{noway}), we arrive at the following conditions:
\be
i Y^\dagger A^T Y = A\,,\quad -iY^\dagger B^T Y = B\,,
\label{fermion-condition1}
\ee
and now the problem translates into finding matrices $Y \in SU(n_f)$ such that these equations have a non-zero solution.

We focus on the case of $n_f = 3$ fermion generations.
First, we immediately deduce from (\ref{fermion-condition1}) that  
\be
\det A = 0\,, \qquad\Tr A^k = 0\,, \quad k=1,2,3\, ,
\ee
and similarly for $B$. 
Next, by performing an appropriate basis change in the fermion space, 
we bring $Y$ to its simplest form:
\be
Y = \mmmatrix{e^{i\beta}}{0}{0}{0}{0}{e^{i\alpha}}{0}{e^{-i\alpha}}{0}\,.\label{Ybasis}
\ee
In this basis, equations (\ref{fermion-condition1}) can be satisfied only 
with the following matrices $A$, $B$, and transformation $Y$:
\bea
\mbox{case 1a}:\ && Y = \mmmatrix{e^{i\beta}}{0}{0}{0}{0}{\pm e^{-i\pi/4}}{0}{\pm e^{i\pi/4}}{0}\,,\quad
A = \mmmatrix{0}{0}{0}{0}{0}{a_{23}}{0}{0}{0}\,,\quad
B = \mmmatrix{0}{0}{0}{0}{0}{0}{0}{b_{32}}{0} \, ;
\label{ABcase1a}\\[2mm]
\mbox{case 1b}:\ && Y = \mmmatrix{e^{i\beta}}{0}{0}{0}{0}{\pm e^{i\pi/4}}{0}{\pm e^{-i\pi/4}}{0}\,,\quad
A = \mmmatrix{0}{0}{0}{0}{0}{0}{0}{a_{32}}{0}\,,\quad
B = \mmmatrix{0}{0}{0}{0}{0}{b_{23}}{0}{0}{0} \, ;
\label{ABcase1b}\\[2mm]
\mbox{case 2}:\ && Y = \mmmatrix{e^{i\beta}}{0}{0}{0}{0}{\pm i}{0}{\mp i}{0},\ 
A = \mmmatrix{0}{a_{12}}{a_{13}}{\mp e^{i\beta} a_{13}}{0}{0}{\pm e^{i\beta} a_{12}}{0}{0},\ 
B = \mmmatrix{0}{b_{12}}{b_{13}}{\pm e^{i\beta} b_{13}}{0}{0}{\mp e^{i\beta} b_{12}}{0}{0}.
\label{ABcase2}
\eea
The two subcases 1a and 1b are related to each other by the permutation of the second and third fermion families. 
In case 2, the $CP$-transformation is also of order 4 in the fermion space, as applying it twice
gives $Y^*Y = \mathrm{diag}(1,-1,-1)$,\footnote{Notice that this minus has nothing to do with
the famous extra minus sign arising in $(CP)^{-2}\psi(x) (CP)^2 = - \psi(x)$. 
Here, we check how $(CP)^2$ acts on the fermion {\em bilinears}, where two such minuses cancel.} 
while in case 1 the $CP$-transformation is in fact of order 8, as $Y^*Y = \mathrm{diag}(1,-i,i)$.
In both cases, the non-zero elements of the Yukawa matrices are exactly those that lead to $CP$-half-odd bilinear combinations.
We will give explicit expressions for these bilinears in the next section.

\section{$CP$-half-odd scalars coupled to fermions in 3HDM}
\label{section:yukawas-ewsb}
\subsection{Yukawa sector before EWSB}

In the previous section, we demonstrated that $CP$-half-odd scalars can in principle couple
to fermions in a $CP$-conserving way via the usual Yukawa interactions, provided the $CP$ acts non-trivially 
not only on scalars but also on fermions.
Now, we want to demonstrate how this coupling arises in the CP4-3HDM,
the model in which $CP$-half-odd scalars were first proposed \cite{IS-2016}.
Notice that in this work we do not attempt to accurately reproduce the experimentally measured values of 
fermion mixing and masses;
we would need to break $CP$ to achieve that.
Here, we just demonstrate that there is no intrinsic inconsistency in this construction.

The Yukawa sector of the model is described as (we only show the quark sector for brevity and use the word fermion generically)
\be
- {\cal L}_Y = 
\bar Q_{Li} \Gamma_{a,ij} d_{Rj} \, \phi_a + \bar Q_{Li} \Delta_{a,ij} u_{Rj} \, \tilde \phi_a + 
\phi_a^\dagger \bar d_{Ri} (\Gamma_{a,ij})^\dagger Q_{Lj} 
+ \tilde \phi_a^\dagger \bar u_{Ri} (\Delta_{a,ij})^\dagger Q_{Lj}\,.\label{SMYukawa} 
\ee
The fermions are chiral, and the left and right fields 
$Q_L$, $d_R$, and $u_R$ can in principle transform differently under the $CP$-transformation, 
with the three matrices $Y_{L}$, $Y_{dR}$, and $Y_{uR}$.
The scalar doublets transform under the $CP$ as $\phi_a \to X_{ab}(\phi_b^\dagger)^T$.
The condition that (\ref{SMYukawa}) is invariant under so-constructed $CP$ transformation is
\be
Y_{L}^\dagger \Gamma_a^* Y_{dR} X^*_{ab} = \Gamma_b\,,\quad 
Y_{L}^\dagger \Delta_a^* Y_{uR} X_{ab} = \Delta_b\,,\label{SMcondition}
\ee
We make the simplifying assumption that $CP$ mixes the left and right fermions in the same way:
\be
Y_{dR} = Y_{uR} = Y_L = Y\,.\label{allY}
\ee
This assumption is natural but not obligatory; we only want to show that even in this case one gets a consistent $CP$-conserving 
Yukawa sector.

Next, we bring the matrix $X$ to the form (\ref{Jb}), and, in this basis, the conditions (\ref{SMcondition}) split into
\bea
&&
Y^\dagger \Gamma_1^* Y = \Gamma_1\,,\quad 
-i Y^\dagger \Gamma_2^* Y = \Gamma_3\,,\quad 
i Y^\dagger \Gamma_3^* Y = \Gamma_2\,,\nonumber\\[1mm]
&&
Y^\dagger \Delta_1^* Y = \Delta_1\,,\quad 
i Y^\dagger \Delta_2^* Y = \Delta_3\,,\quad 
-i Y^\dagger \Delta_3^* Y = \Delta_2\,.\label{SMconditions2}
\eea
Then we make the basis change
in the fermion space which brings $Y$ to the form (\ref{Ybasis}).
In this basis, we again find two cases for non-trivial solutions for $\Gamma_a$ and $\Delta_a$:
\begin{itemize}
\item
{\bf case 1}: $\alpha = \pm\pi/4 + \pi k$:
\be
\Gamma_1 = \mmmatrix{g_1}{0}{0}{0}{g_2}{0}{0}{0}{g_2^*}\,,\quad
\Gamma_2 = \mmmatrix{0}{0}{0}{0}{0}{g_{23}}{0}{g_{32}}{0}\,,\quad
\Gamma_3 = \mmmatrix{0}{0}{0}{0}{0}{\pm g_{32}^*}{0}{\mp g_{23}^*}{0}\,.
\label{SM-case1}
\ee
\item
{\bf case 2}: $\alpha = \pm \pi/2$:
\be
\Gamma_1 = \mmmatrix{g_1}{0}{0}{0}{g_2}{g_3}{0}{-g_3^*}{g_2^*}\,,\quad
\Gamma_2 = \mmmatrix{0}{g_{12}}{g_{13}}{g_{21}}{0}{0}{g_{31}}{0}{0}\,,\quad
\Gamma_3 = \pm \mmmatrix{0}{- e^{-i\beta}g_{13}^*}{ e^{-i\beta} g_{12}^*}{ e^{i\beta}g_{31}^*}{0}{0}{- e^{i\beta}g_{21}^*}{0}{0}\,.
\label{SM-case2}
\ee
\end{itemize}
In both cases $g_1$ is real and all other entries are complex and independent.
The expressions for $\Delta_a$ are of the same form, with parameters $d_i$ instead of $g_i$
and with the exchange of index $2 \leftrightarrow 3$. 
Thus, we have constructed the desired $CP$-conserving Yukawa sector based on the order-4 $CP$-symmetry.

\subsection{Yukawa sector after EWSB}

To keep the order-4 $CP$-symmetry after the electroweak symmetry breaking, 
we select the vacuum expectation value alignment $v_i = (v,\,0,\,0)$. This choice is symmetry-protected and technically natural,
and it arises in a significant part of the entire scalar potential parameter space.

For the sake of illustration, we turn to (and focus only on) charged leptons. 
We use the familiar notation $e$, $\mu$, and $\tau$, to label fermion generations,
but we do not mean that they have the true properties of the charged leptons observed in experiment.
We reiterate that in this paper we only explore the internal consistency of the construction;
whether a more elaborate model with explicit or spontaneous violation of the order-4 $CP$ 
can accurately describe fermion properties is left for a subsequent publication.

The charged lepton Yukawa lagrangian is 
\be
- {\cal L}_Y = 
\bar \ell_{Li} \Gamma_{a,ij} \ell_{Rj} \, \phi_a + 
\bar \ell_{Ri} (\Gamma_{a,ij})^\dagger \ell_{Lj} \, \phi_a^*\,.\label{SMleptons} 
\ee
Here and below, $\phi_a$ always stands for the neutral components of the doublets:  $\phi_a \equiv \phi_a^0$.
Since the $CP$-symmetry mixes the second and third generations, 
they must be mass-degenerate. Indeed, the masses come from $\Gamma_1$ and
are equal to $m_e^2 = |g_1|^2v^2/2$ and $m_\mu^2 = m_\tau^2 = |g_2|^2v^2/2$ (case 1)
or $m_\mu^2 = m_\tau^2 = (|g_2|^2 + |g_3|^2) v^2/2$ (case 2). Notice that in case 2,
when switching to the $\Gamma_1$-diagonal basis, $\Gamma_{2,3}$ have the same form as in (\ref{SM-case2}) 
just with redefined parameters $g_{ij}$.

The SM-like Higgs boson from the first doublet couples to the fermions exactly as in the SM. 
The neutral Higgses from the second and third doublets induce non-diagonal interactions.
Let us start with case 1 given in (\ref{SM-case1}) with $\alpha = -\pi/4$.
Written explicitly, the Yukawa interactions with neutral scalars are
\bea
-{\cal L}_Y&=&\bar\mu_L (g_{23} \phi_2 - g_{32}^*\phi_3)\tau_R + \bar\tau_L(g_{32}\phi_2 + g_{23}^*\phi_3)\mu_R \nonumber\\
&+&
\bar\tau_R (g^*_{23} \phi_2^* - g_{32}\phi_3^*)\mu_L 
+ \bar\mu_R(g_{32}^*\phi_2^*+ g_{23}\phi_3^*)\tau_L\label{leptons-case1}\\[1mm]
&=&
g_{23}\left(\bar\mu_L\tau_R \phi_2 + \bar\mu_R\tau_L\phi_3^*\right) 
+ g_{32}\left(\bar \tau_L \mu_R \phi_2 - \bar\tau_R \mu_L \phi_3^*\right)\nonumber\\
&+& g_{23}^*\left(\bar\tau_R \mu_L\phi_2^* + \bar\tau_L\mu_R\phi_3\right) 
+ g_{32}^*\left(\bar \mu_R \tau_L \phi_2^* - \bar\mu_L \tau_R \phi_3\right)\,.\label{leptons2-case2}
\eea
The last form exposes the remaining order-4 $CP$-symmetry. For example, 
\bea
\bar\mu_L\tau_R \toCP -i\bar\mu_R\tau_L\,,\quad \bar\mu_R\tau_L \toCP -i\bar\mu_L\tau_R\,,\label{bilinears-leptons1}
\eea
which compensates the $\phi_2 \toCP i \phi_3^*$ and $\phi_3^*\toCP i \phi_2$ transformation. 
One can further combine bilinears into $CP$-eigenstates:
\be
\bar\mu\tau \toCP -i \bar\mu\tau\,,\quad \bar\tau\mu \toCP i \bar\tau\mu\,,\quad
\bar\mu\gamma_5\tau \toCP i \bar\mu\gamma_5\tau\,,\quad \bar\tau\gamma_5\mu \toCP -i \bar\tau\gamma_5\mu\,.\label{bilinears-CP-states}
\ee
Remarkably, the fermion bilinears shown here are $CP$-half-odd, with quantum number $q=\pm1$.
It is also remarkable that insertion of $\gamma_5$ changes $q$ by two units, 
which is equivalent of an extra $CP$-oddness, just like it happens in the usual case.

Finally, we switch from $\phi_2^0$, $\phi_3^0$ to the $CP$-half-odd scalars $\varPhi$, $\varphi$ as shown
in Appendix~\ref{appendix:scalar-CP4-3HDM}. 
Then, the final form for the Yukawa interactions between $CP$-half-odd scalar and the fermions is
\be
-{\cal L}_Y = (\bar\mu\tau) (g \varPhi - \tilde g \varphi) + (\bar\tau\gamma_5\mu) (\tilde g^* \varPhi + g^* \varphi) + h.c.,\label{final-case1}
\ee
where 
\be
g = {c_\gamma g_{23} - s_\gamma g_{32}^* \over \sqrt{2}}\,,\quad 
\tilde g = {s_\gamma g_{23} + c_\gamma g_{32}^*\over \sqrt{2}}\,,\quad \tan 2\gamma = - \lambda_6/\lambda_5\,.
\ee
This interaction is exactly of the type (\ref{ABcase1a}) for both $CP$-half-odd fields.

For case 2, the Yukawa interactions can be grouped as
\bea
-{\cal L}_Y&=&g_{12}(\bar e_L\mu_R\phi_2 + \bar \tau_R e_L \phi_3^*) 
+ g_{13}(\bar e_L \tau_R \phi_2 - \bar \mu_R e_L \phi_3^*)\nonumber\\
&+& g_{21}(\bar \mu_L e_R\phi_2 - \bar e_R \tau_L \phi_3^*) 
+ g_{31}(\bar \tau_L e_R \phi_2 + \bar e_R \mu_L \phi_3^*)
+ h.c.\label{leptons-case2}
\eea
Again, we can group bilinears into $CP$-eigenstates, for example,
\be
\bar e \mu + \bar \tau e \toCP -i (\bar e \mu + \bar \tau e)\,,\quad
\bar e \gamma_5 \mu + \bar \tau \gamma_5 e \toCP i (\bar e \gamma_5 \mu + \bar \tau \gamma_5 e)\,.
\ee
Like in case 1, these bilinears are $CP$-half-odd, and insertion of $\gamma_5$ introduces an extra $CP$-oddness.
Finally, switching to the $CP$-half-odd scalars 
\bea
-{\cal L}_Y &=& (\bar e \mu + \bar\tau e)(g_+\varPhi - \tilde g_+ \varphi) - 
(\bar \mu e - \bar e\tau)(\tilde g_-^* \varPhi + g_-^*\varphi)\nonumber\\
&+& (\bar e \gamma_5 \mu - \bar\tau \gamma_5 e)(g_-\varPhi - \tilde g_- \varphi) 
- (\bar \mu \gamma_5 e + \bar e \gamma_5 \tau)(\tilde g_+^* \varPhi + g_+^*\varphi) + h.c.,\label{final-case2}
\eea
where we introduced the combined couplings
\bea
&&g_+ = {(g_{12} + g_{31})c_\gamma - (g_{13}^*+g_{21}^*)s_\gamma \over 2\sqrt{2}}\,,\quad 
\tilde g_+ = {(g_{12}+g_{31})s_\gamma + (g_{13}^*+g_{21}^*)c_\gamma \over 2\sqrt{2}}\,,\quad \nonumber\\
&&g_- = {(g_{12} - g_{31})c_\gamma - (g_{13}^* -g_{21}^*)s_\gamma \over 2\sqrt{2}}\,,\quad 
\tilde g_- = {(g_{12} - g_{31})s_\gamma + (g_{13}^*-g_{21}^*)c_\gamma \over 2\sqrt{2}}\,,
\eea
all of them being independent.
These interaction terms are exactly what is encoded in (\ref{ABcase2}).

\subsection{Discussion}

The resulting Yukawa interactions (\ref{final-case1}), (\ref{final-case2}) 
exhibit a peculiar asymmetric pattern of couplings of the $CP$-half-odd scalars and their conjugates to fermion pairs.
It is tempting to interpret interaction terms such as $\bar \mu \tau \varPhi$
as a source of lepton flavour violation. 

However when reading physical processes off such interactions, one must not forget that,
according to the convention adopted, the single-fermion particle and antiparticle states
are linked via the conserved generalized $CP$ transformation in the fermion space.
As a result, fermion and its antifermion creation operators belong to {\em different} fields, 
just as it was the case for scalars, see discussion after Eq.~(\ref{phi1phi2CP3}). 
Therefore, the interaction $\bar \mu \tau \varPhi$
in case 1 describes the $\varPhi$ decay to a $\mu^+\mu^-$ pair (or $\tau^+\tau^-$ transition into $\varPhi^*$), 
while $\bar\tau\mu \varPhi^*$ describes the $\varPhi^*$ decay to a $\tau^+\tau^-$ pair.
As a result, $\varPhi$ and $\varPhi^*$ have different decay preferences, 
but since they are not antiparticles of each other, these results are hardly surprising.
The situation is less trivial in case 2, where at least the lepton-flavor-violating coupling between
$e$ and $\mu/\tau$ exists.

Still, one might not be fully satisfied with our convention of identifying the particle and antiparticle states
for fermions. The fermions are charged and participate in the electromagnetic interactions
via the standard interaction terms $\bar \ell_i \gamma^\mu \ell_i A_\mu$ that are diagonal in fermion flavor.
Expressing them in terms of creation and annihilation operators,
one sees that they correspond not only to subprocess $\mu^- \to \mu^- \gamma$ but also to 
$\mu^- \tau^+ \to \gamma$. One is lead to the counter-intuitive conclusion that despite the fact that a fermion can emit a photon
without changing its flavor, it must pick up a different fermion to annihilate into a single photon.

One can revert the fermion-antifermion convention back to the usual one, in which a single fermion field
contains the creation operator of a particle and the annihilation operator of its antiparticle.
In this, more physically appealing case, fermion annihilates together with its antifermion.
However, in this case the Yukawa interactions (\ref{final-case1}), (\ref{final-case2}) will be manifestly
$CP$-violating, despite the fact that $CP$ is conserved in this model by all commonly accepted standards.

To summarize this discussion, our model reveals a surprising clash between two different conventions
for particle-antiparticle assignments for charged fermions.
One is ``technical'', it is consistent with the conserved $CP$-symmetry, but it leads to counter-intuitive 
transitions like $\mu^- \tau^+ \to \gamma$. The other is ``physical''; it requires
that at tree-level particles can only annihilate with their antiparticles into a photon.
But in this case one must accept that a $CP$-conserving model leads to manifest $CP$-violation.

There is a third way: to simply avoid assigning who is antiparticle of whom.
In this case, there is no such transformation as $C$-parity, and the $CP$-symmetry of order 4
the model possesses is just a peculiar symmetry linking different fields.
However it is not clear how one should phrase the physical phenomenon of $CP$ violation
and baryogenesis within this ``$C$-agnostic'' point of view.

Yet another possibility is that it is premature to draw any phenomenological conclusion from 
the above observations because this model features not only a conserved $CP$ but also the mass-degenerate
$\mu$ and $\tau$. 
It will be interesting to see whether in a phenomenologically relevant version of CP4-3HDM
with a spontaneously broken $CP$ any interaction of this type persists and leads to observable signals.

To this end, we note that our model bears similarity with two versions of 2HDM studied recently
in \cite{Maniatis:2007de,Maniatis:2009vp}, dubbed the 2HDM with ``maximal $CP$-symmetry'', 
and in \cite{Ferreira:2010bm}.
In both cases, one imposes a higher-order GCP on the Higgs potential and
then extends the symmetry to the Yukawa sector, allowing for mixing between fermion families.
The first \cite{Maniatis:2007de,Maniatis:2009vp}
exploits essentially the same order-4 symmetry transformation, 
but since it is applied to two doublets, it effectively becomes an order-2 symmetry.
Indeed, applying it twice leads to the overall minus sign in the scalar sector which can be removed by the global sign flip.
The Yukawa sector turns out to be very restrictive, and upon symmetry breaking, 
leads to one massless fermion generation and to strong lepton flavor violation.
The second work \cite{Ferreira:2010bm} asked which higher-order GCP can be imposed
on 2HDM without running into immediate troubles with the quark sector. That work
also confirmed that an order-4 transformation 
would lead to one massless fermion generation and thus was considered unphysical,
but another GCP transformation with rotation angle $\alpha=\pi/3$ turned out compatible 
with the experimentally measured quark masses and mixing pattern. 
We notice in passing that the origin of this special value lies in one additional discrete abelian symmetry group
$\Z_3$ with exists in 2HDM Yukawa sector. Thus, imposing GCP of order 6 is equivalent
to imposing a usual $CP$ and an order-3 family symmetry transformation. 

Our CP4-3HDM differs in several important ways from those two models.
First, the presence of an additional doublet renders the symmetry genuine order-4, not order-2 transformation.
Second, the third doublet can acquire the vev after EWSB making it possible to 
keep CP4 unbroken. It is this residual symmetry that allows us to identify
the $CP$-half-odd neutral scalars.
Third, when CP4 is spontaneously broken, which can easily happen in a larger part of the parameter space,
the resulting fermion mass matrices do not lead to massless fermions. This is again due to the presence
of a third doublet with different Yukawa matrices.
Therefore, unlike 2HDM, this model may lead to a phenomenologically relevant fermion sector
with interesting family-violating signatures. Moreover, these signatures
do not have to be dramatic because they beat against the SM-like Yukawa structure.
Building and exploring a CP4-3HDM with realistic fermion sector is the next step in exploration of this model,
and we delegate this task to a future study.

Finally, we briefly comment on possibility of $CP$-symmetries of even higher order.
First, we mention the basic group-theoretic fact that, if $p_1$ and $p_2$ are two distinct primes,
then $\Z_{p_1p_2} \simeq \Z_{p_1} \times \Z_{p_2}$. Therefore, if for example the $CP$-symmetry is of order six,
then the symmetry group can be factored into the usual $CP$-symmetry and a family symmetry group $\Z_3$.
The only case when the $CP$-symmetry of order $p$ cannot be factored into a smaller-order GCP
and a family symmetry is when $p = 2^k$. Thus, $CP$-symmetries of order $8, 16, \dots$ are in principle possible.

Explicitly constructing a model with order-8 GCP (and higher) and no other accidental symmetries is a separate task.
If located purely in the scalar sector of multi-Higgs-doublet models, 
it must involve more than three Higgs doublets; this is because all abelian symmetry groups 
of 3HDM were listed in \cite{abelian} and no such example was found. 
One would need to repeat this procedure for 4HDM to see if there is such a model.
The fact that the renormalizable potential only has quadratic and quartic terms does not contradict
this possibility. Most likely, such as model will contain new complex scalars $\varPhi_1$ and $\varPhi_3$
with $CP$-charges $q=1$ and $q=3$, all defined modulo 8 that would interact 
via quartic interactions $\varPhi_1^2 \varPhi_3^2$.
It would be interesting to see a specific model realizing this idea.

\section{Conclusions}

In this paper we further explored the origin and properties of the peculiar $CP$-half-odd scalars 
recently found in the CP4-3HDM, a 3HDM model based on an order-4 $CP$ symmetry \cite{IS-2016}.
The central theme is the unavoidable ambiguity --- and consequently an enhanced freedom --- 
of defining the discrete transformations $C$, $P$ and $CP$ in models with several zero-charge fields.
In this situation, the class of physically acceptable $CP$ transformations is broader
than the traditionally appreciated generalized $CP$.
In fact, the absence of gauge charges blurs the distinction between particles and antiparticles to such an extent
that the same $CP$ transformation can resemble a $P$ transformation through a mere basis change.
Although some previous publications hinted at this formal possibility, no specific example of such a construction was known.
We found and explored such examples.

In order to accompany the reader through the meander of subtleties, we gave in this paper 
a pedagogical presentation, through examples, of the salient features for $C$, $P$ and $CP$ symmetries acting on scalars.
We also linked some of our material to results obtained by others in different approaches to $CP$ symmetries.

In the second part of the paper, we showed that $CP$-half-odd scalars can be coupled to fermions
via the usual Yukawa interactions in the $CP$-conserving way, provided the $CP$ acts on fermions
as a family-mixing generalized $CP$ transformation.
We found two classes of Yukawa matrices for the case of three fermion generations.
Phenomenologically, it implies that the $CP$-half-odd scalars introduced in CP4-3HDM 
do not have to be inert after all.

The purpose of this work was to show the internal consistency of $CP$-half-odd scalars
and of their Yukawa interactions. To this end, when analyzing CP4-3HDM, we deliberately selected
the vacuum alignment which conserves the order-4 $CP$ symmetry.
Certainly, the model with the exact $CP$ symmetry cannot reproduce the experimentally
observed fermion masses and mixing. Given the results obtained in this work, 
one is now led to ask whether 
a similar model, based on the spontaneous or explicit breaking of the order-4 $CP$ symmetry,
can accurately reproduce the flavor sector and whether it will be more economical 
than other model-building attempts.
This investigation is delegated to a future publication. 

{\bf Acknowledgements.} We are thankful to 
Bhupal Dev, Howard Haber, Celso Nishi, Leonardo Pedro, Jo\~{a}o Silva, Andreas Trautner,
and the anonymous referee
for useful comments and discussions.
The work of I.P.I. was supported by the Portuguese
\textit{Fun\-da\-\c{c}\~{a}o para a Ci\^{e}ncia e a Tecnologia} (FCT)
through the Investigator contract IF/00989/2014/CP1214/CT0004
under the IF2014 Program and in part
by contracts UID/FIS/00777/2013 and CERN/FIS-NUC/0010/2015,
which are partially funded through POCTI, COMPETE, QREN, 
and the European Union. 
The support from CONACYT project CB­-2015-­01/257655 (M\'exico) is also acknowledged
by A.A. and E.J.

\appendix
\section{Scalar sector of the CP4-3HDM}
\label{appendix:scalar-CP4-3HDM}

The Higgs potential in the CP4-3HDM considered in \cite{IS-2016}
is $V=V_0+V_1$, where
\bea
V_0 &=& - m_{11}^2 (\phi_1^\dagger \phi_1) - m_{22}^2 (\phi_2^\dagger \phi_2 + \phi_3^\dagger \phi_3)
+ \lambda_1 (\phi_1^\dagger \phi_1)^2 + \lambda_2 \left[(\phi_2^\dagger \phi_2)^2 + (\phi_3^\dagger \phi_3)^2\right]
\nonumber\\
&+& \lambda_3 (\phi_1^\dagger \phi_1) (\phi_2^\dagger \phi_2 + \phi_3^\dagger \phi_3)
+ \lambda'_3 (\phi_2^\dagger \phi_2) (\phi_3^\dagger \phi_3)\nonumber\\
&+& \lambda_4 \left[(\phi_1^\dagger \phi_2)(\phi_2^\dagger \phi_1) + (\phi_1^\dagger \phi_3)(\phi_3^\dagger \phi_1)\right]
+ \lambda'_4 (\phi_2^\dagger \phi_3)(\phi_3^\dagger \phi_2)\,,\label{V0}
\eea
with all parameters being real, and
\be
V_1 = \lambda_5 (\phi_3^\dagger\phi_1)(\phi_2^\dagger\phi_1)
+ {\lambda_6 \over 2}\left[(\phi_2^\dagger\phi_1)^2 - (\phi_1^\dagger\phi_3)^2\right] +
\lambda_8(\phi_2^\dagger \phi_3)^2 + \lambda_9(\phi_2^\dagger\phi_3)(\phi_2^\dagger\phi_2-\phi_3^\dagger\phi_3) + h.c.
\label{V1b}
\ee
with real $\lambda_5$, $\lambda_6$, and complex $\lambda_8$, $\lambda_9$.
This potential is invariant under the generalized $CP$ transformation $J_X$ defined in (\ref{GCP}), with
\be
X =  \left(\begin{array}{ccc}
1 & 0 & 0 \\
0 & 0 & i  \\
0 & -i & 0
\end{array}\right)\,.
\label{Jb}
\ee
A key observation is that $J_X$ is an order-4 transformation:
\be
J_X^2 = X X^* = \mathrm{diag}(1,\,-1,\,-1) \not = \mathbb{I}\,, \quad J_X^4 = \mathbb{I} \equiv \mathrm{diag}(1,\,1,\,1) \,.
\ee
For generic values of the coefficients, this potential has no other Higgs-family or $CP$-symmetries
apart from powers of $J_X$ \cite{abelian}.
Eqs.~(\ref{V0}) and (\ref{V1b}) define the most general renormalizable potential to which one arrives
starting from any 3HDM invariant under an order-4 $CP$ and applying basis change transformations
to reduce the number of complex coefficients.

Next, we select the $CP$-conserving vacuum alignment: $\lr{\phi_1^0} = v/\sqrt{2}$, $\lr{\phi_2} = \lr{\phi_3} = 0$.
For physical scalars, we get the SM-like Higgs with mass
$m_{h_{SM}}^2 = 2\lambda_1 v^2 = 2m_{11}^2$,
and a pair of degenerate charged Higgses with
$m_{H^\pm}^2 = \lambda_3 v^2/2 - m_{22}^2$.
In the neutral scalar sector, the mass matrices for real $h_{2,3}$ and imaginary $a_{2,3}$ components of $\phi_{2,3}^0$ split,
\bea
&&M_{h_2,\,h_3} = \mmatrix{a + b}{c}{c}{a - b}\,,\quad
M_{a_2,\,a_3} = \mmatrix{a - b}{-c}{-c}{a + b}\,,\nonumber\\
&&a = {1 \over 2}v^2 (\lambda_3+\lambda_4) - m_{22}^2 = m_{H^\pm}^2 + {1\over 2}v^2\lambda_4\,,
\quad b = {1 \over 2}v^2\lambda_6\,,\quad c = {1\over 2}v^2\lambda_5\,,
\eea
and lead to the same physical scalar spectrum in both spaces:
\be
M^2 = a + \sqrt{b^2 + c^2}\,,\quad m^2 = a - \sqrt{b^2 + c^2}\,.
\ee
The diagonalization of both mass matrices is performed by a rotation with the angle $\alpha$ defined as
$\tan2\alpha = \lambda_5/\lambda_6$,
but it proceeds in the opposite directions for $h$'s and $a$'s.
The two heavier scalars $H, A$ and the two ligher scalars $h$ and $a$ are related to initial fields as
\be
\doublet{H}{h} = \mmatrix{c_\alpha}{s_\alpha}{-s_\alpha}{c_\alpha}\doublet{h_2}{h_3}\,,\quad
\doublet{a}{A} = \mmatrix{c_\alpha}{s_\alpha}{-s_\alpha}{c_\alpha}\doublet{a_2}{a_3}\,.\label{rotations}
\ee
Note that, upon (\ref{rotations}), $\phi_2^0$ and $\phi_3^0$ transform to
\be
c_\alpha \phi_2^0 + s_\alpha\phi_3^0 = {1 \over\sqrt{2}}(H + i a)\,,
\quad
-s_\alpha \phi_2^0 + c_\alpha\phi_3^0 = {1 \over\sqrt{2}}(h + i A)\,.
\ee
The real neutral fields $H, A, h, a$ are not $CP$-eigenstates:
\be
H \toCP A\,, \quad A\toCP -H\,, \quad h \toCP -a\,, \quad a\toCP h\,.
\ee
One can combine them into neutral fields,
$\varPhi = (H - i A)/\sqrt{2}$, $\varphi = (h + i a)/\sqrt{2}$,
which {\em are} $CP$ and mass eigenstates:
\be
\varPhi \toCP i\varPhi\,,\quad \varphi \toCP i\varphi\,.\label{J-eigenstates}
\ee
One can then quantify the $CP$ properties with the global quantum number $q$ defined modulo 4,
and assign $q=+1$ to $\varPhi$, $\varphi$, and $q=-1$ to their conjugate fields.
All other neutral fields are either $CP$-odd, $q=+2$, or $CP$-even, $q=0$.
This quantum number can also be associated with single-particle states as defined in section~\ref{section-CP4}.
Since $CP$ is a good symmetry of the lagrangian and of the vacuum, it commutes with the hamiltonian.
Therefore, in any transition between initial and final states with definite $q$, this quantum number is conserved.

\end{document}